\documentclass[aps,prb,twocolumn,groupedaddress,showpacs]{revtex4}

\usepackage{graphicx}

\begin{document}


\title{Molecular orientational dynamics of the endohedral fullerene Sc$_{3}$N@C$_{80}$ as probed by $^{13}$C and $^{45}$Sc NMR}


\author{K.~R.~G\'{o}rny}
\author{C.~H.~Pennington}
\author{J.~A.~Martindale}
\thanks{Author to whom correspondence should be addressed.  E-mail address is martindale.5@osu.edu.}
\affiliation{Department of Physics, The Ohio State University, 191 West Woodruff Avenue, Columbus, Ohio 43210}

\author{J.~P.~Phillips}
\author{S.~Stevenson}
\affiliation{Department of Chemistry and Biochemistry, University of Southern Mississippi, 118 College Drive \#5043, Hattiesburg, Mississippi 39406}

\author{I.~Heinmaa}
\author{R.~Stern}
\affiliation{National Institute of Chemical Physics and Biophysics, Akadeemia tee 23, 12618 Tallinn, Estonia}


\date{\today}

\begin{abstract}
We measure $^{13}$C and $^{45}$Sc NMR lineshapes and spin-lattice relaxation times (T$_{1}$) to probe the orientational dynamics of the endohedral metallofullerene Sc$_{3}$N@C$_{80}$.  The measurements show an activated behavior for molecular reorientations over the full temperature range with a similar behavior for the temperature dependence of the $^{13}$C and $^{45}$Sc data.  Combined with spectral data from Magic Angle Spinning (MAS) NMR, the measurements can be interpreted to mean the motion of the encapsulated Sc$_{3}$N molecule is independent of that of the C$_{80}$ cage, although this requires the similar temperature dependence of the $^{13}$C and $^{45}$Sc spin-lattice relaxation times to be coincidental.  For the Sc$_{3}$N to be fixed to the C$_{80}$ cage, one must overcome the symmetry breaking effect this has on the Sc$_{3}$N@C$_{80}$ system since this would result in more than the observed two $^{13}$C lines.
\end{abstract}


\maketitle

\section{\label{Introduction}Introduction}
The enclosure of molecules within the carbon cages of fullerenes promises to add yet more richness to the array of possibilities associated with fullerene chemistry.  To date, a limiting factor in the investigations of such materials has been the inability to synthesize them in macroscopic quantities.  Exceptions to this rule include molecules made using the Òtrimetallic nitride templateÓ (TNT) synthesis method,\cite{steve} which can produce tens of milligrams of material of the form M$_{3}$N@C$_{80}$, where M is a metal atom (including Sc, Y, and many Lanthanides).  The C$_{80}$ molecular cage enclosing the M$_{3}$N complex has icosahedral symmetry and is stable only because of charge transfer from the nitride cluster within.

Experiments on the prototypical fullerene C$_{60}$ have revealed orientational dynamics of the molecules in the solid state.  For example, x-ray diffraction and calorimetry showed a transition from orientational disorder to order around 250 K.\cite{heiney}  NMR measurements confirmed a sharp change in behavior at this temperature but continued to indicate rapid molecular reorientations even at temperatures below this transition.\cite{tycko,johnson,review}  NMR investigators concluded that a snapshot of the sub-250 K structure would reveal orientational order, but the molecules would make discrete orientational ``jumps'' with a characteristic correlation time dependent on temperature. This was named the ``ratchet'' phase, and the high-temperature phase was called the ``rotator'' phase.  One might expect such phases to exist in M$_{3}$N@C$_{80}$ as well because the C$_{80}$ cage has icosahedral symmetry identical to that of C$_{60}$.

We report results of $^{13}$C and $^{45}$Sc NMR as probes of the molecular orientational dynamics of Sc$_{3}$N@C$_{80}$ and compare and contrast with C$_{60}$.  Our measurements reveal no sign of the rotator/ratchet transition found in C$_{60}$ but instead show an activated behavior for molecular reorientations over the full temperature range from 20 K -- 330 K.  The spin-lattice relaxation times (T$_{1}$) for both $^{13}$C and $^{45}$Sc display similar temperature behavior, with a minimum between 100 K and 200 K, suggesting a similar physical cause for their motion.  We also report Magic Angle Spinning (MAS) NMR spectra for $^{13}$C at lower temperature that show only two carbon lines (proving the icosahedral symmetry of the C$_{80}$ cage), although the lines broaden somewhat as the temperature decreases which may be indicative of a loss of spherical symmetry.  Taking all of these data together along with previous experimental and theoretical results, one may conclude that the endohedral Sc$_{3}$N is not fixed to the C$_{80}$ cage on the time scale of the NMR experiments, although this would require the similar temperature dependence for the $^{13}$C and $^{45}$Sc spin-lattice relaxation times to be a coincidence.  To argue that the Sc$_{3}$N is fixed to the C$_{80}$ cage, one must overcome the fact that this breaks the spherical symmetry of the cage and therefore should produce more carbon lines than just the two we observe.

\section{\label{Expt}Experimental}
The procedure for preparation of Sc$_{3}$N@C$_{80}$ by the group at the University of Southern Mississippi\cite{steve} and the MAS NMR apparatus of the Tallinn group\cite{stern} have been described elsewhere.  The non-MAS NMR measurements at The Ohio State University were carried out at a magnetic field strength of 8.8 T utilizing a home-built spectrometer.  Spectra for $^{13}$C and $^{45}$Sc were obtained via Fourier Transform of spin-echo signals.  The spin-lattice relaxation times were measured with the saturation recovery technique utilizing 200 pulses to obliterate the magnetization.

\section{\label{Results}Results}
\begin{figure}
\includegraphics[width=3in,height=3in,trim=0 0 0 0]{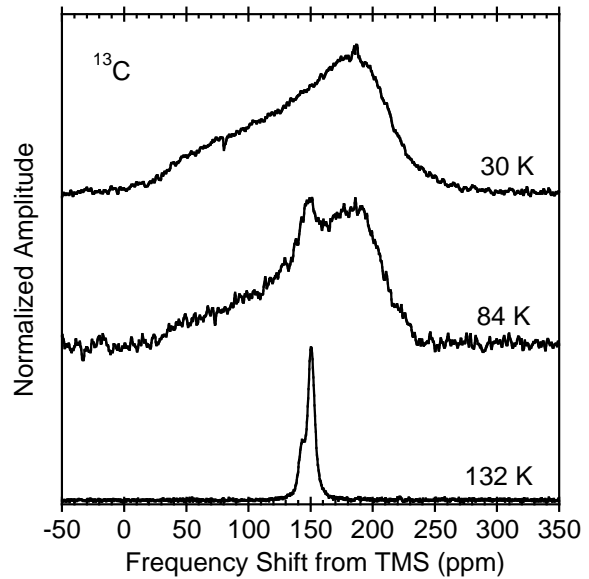}
\caption{\label{carbonline} $^{13}$C spectra for Sc$_{3}$N@C$_{80}$ show molecular reorientation of the C$_{80}$ cages via motional narrowing.  All data were acquired at applied field 8.8 Tesla with line position measured in parts per million (ppm) with respect to tetramethyl silane (TMS).  At low temperature (30 K) a chemical shift anisotropy (CSA) powder pattern is observed with features indicating principal values of the chemical shift tensor of $\sigma_{11}  \approx$ 220 ppm, $\sigma_{22} \approx$ 180 ppm, and $\sigma_{33} \approx$ 30 ppm.  At high temperature (represented by the data for T = 132 K), the CSA powder pattern becomes motionally narrowed due to rapid reorientational motion of the C$_{80}$ cage.  The narrowed lineshape shows a peak at 155 ppm and a distinct shoulder at 148 ppm, in agreement with solution NMR data for the primary isomer of Sc$_{3}$N@C$_{80}$ (although our results are shifted to higher frequency by $\approx$ 10 ppm).  At intermediate temperature (T = 84 K), we observe a superposition of the high- and low-temperature spectra, indicating the motion of the cages has begun to slow.}
\end{figure}

Spectra for $^{13}$C (shown in Fig.~\ref{carbonline} for three different temperatures) reveal dynamical behavior of the C$_{80}$: at high temperature, the cages undergo rapid reorientational motion that causes motional narrowing of the line, and this motion freezes out as the temperature is lowered.  In the high-temperature regime (T = 132 K in Fig.~\ref{carbonline}), the spectrum shows a peak and a distinct shoulder.  A fit at this temperature yields two lines at 155 ppm and 147 ppm with intensities in the ratio of 3:1.  These data agree with previous solution NMR results,\cite{steve} although the line positions reported here are approximately 10 ppm higher than in Ref.~\onlinecite{steve}.  Part of this discrepancy results from a slight temperature dependence to the line positions (see MAS data below); as to the remainder, we have no definitive explanation but speculate a C$_{80}$-solvent interaction in the solution NMR might be responsible.  Moving to the low temperature region (T = 30 K in Fig.~\ref{carbonline}), we find a chemical shift anisotropy (CSA) powder pattern with features indicating principal values of the chemical shift tensor of $\sigma_{11}  \approx$ 220 ppm, $\sigma_{22} \approx$ 180 ppm, and $\sigma_{33} \approx$ 30 ppm (measured relative to TMS = tetramethyl silane), values not significantly different from those reported for C$_{60}$.\cite{tycko,yannoni}  At intermediate temperatures (T = 84 K in Fig.~\ref{carbonline}), a superposition of the high- and low-temperature spectra occurs.

\begin{figure}
\includegraphics[width=3in,height=3in,trim=0 0 0 0]{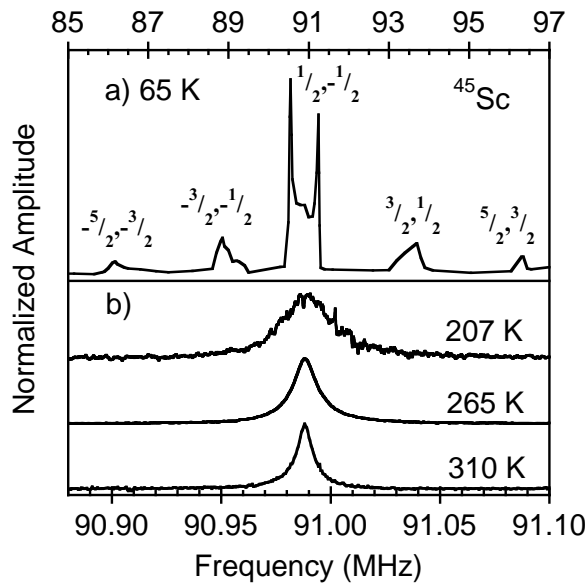}
\caption{\label{scandiumline} $^{45}$Sc spectra for Sc$_{3}$N@C$_{80}$ show reorientation of the Sc$_{3}$N molecule via motional narrowing.  All data were measured at applied field 8.8 Tesla.  The horizontal scale for (a) is shown at the top while the scale for (b) is at the bottom; both show frequency in units of MHz.  The results are similar to those for $^{13}$C in Fig.~\ref{carbonline}:  a powder pattern at low temperature (a) and motional narrowing at high temperature (b).  In (a), rather than showing a chemical shift anisotropy powder pattern as was the case for $^{13}$C, the low-temperature data for the spin-7/2 $^{45}$Sc nucleus reveal a full first order quadrupole powder pattern.  The various transitions are labeled in the figure; the outermost satellites were not observed.  This spectrum is characterized by asymmetry parameter $\eta \approx$ 0 and electric field gradient parameter e$^{2}$qQ/h $\approx$ 67.9 MHz.  At higher temperatures such as those shown in (b), the data demonstrate that the first order powder pattern is motionally narrowed, and only one peak is observed.}
\end{figure}

The $^{45}$Sc spectral data (Fig.~\ref{scandiumline}) show the same motional narrowing behavior as do the $^{13}$C data.  The spectrum for $^{45}$Sc (spin 7/2) is richer than that of $^{13}$C (spin 1/2) because $^{45}$Sc has a nuclear electric quadrupole moment Q = $-$0.22 x 10$^{-24}$ cm$^{2}$ which interacts with the local electric field gradient (EFG) at the nuclear site.  At low temperature (T = 65 K) we observed a full first order quadrupole powder pattern\cite{cps,cohenreif} characterized by asymmetry parameter $\eta \approx$ 0 and electric field gradient parameter e$^{2}$qQ/h $\approx$ 67.9 MHz.  At higher temperatures the powder pattern is motionally narrowed, and only one peak is observed with linewidth of order tens of kHz.  This motional narrowing demonstrates molecular reorientation on a time scale shorter than 0.1 $\mu$s.

\begin{figure}
\includegraphics[width=3in,height=3in,trim=0 0 0 0]{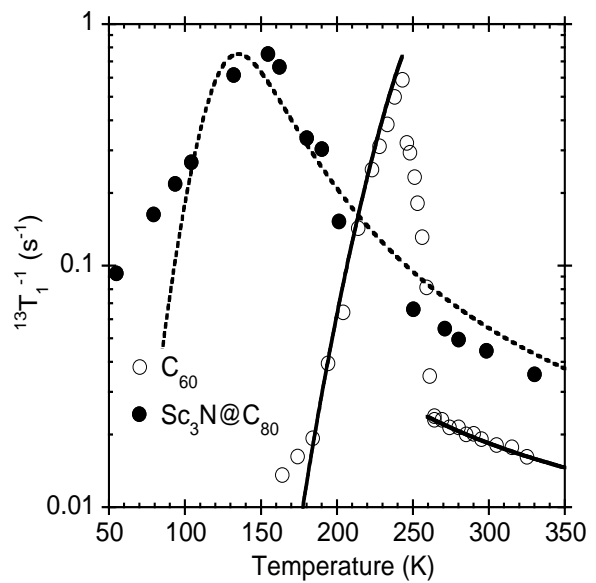}
\caption{\label{carbonrate} $^{13}$C spin-lattice relaxation rates 1/T$_{1}$ vs. temperature reveal a T$_{1}$ minimum for both Sc$_{3}$N@C$_{80}$ (solid circles) and C$_{60}$ (open circles), indicating molecular reorientations are occurring in both systems.  The C$_{60}$ data come from Tycko \textit{et al.}\cite{tycko} The lines are fits to the data using the model of Tycko \textit{et al.} described in the text that ascribes the relaxation to modulations of the chemical shift due to molecular reorientation.  The solid lines apply to C$_{60}$ and the dashed line to Sc$_{3}$N@C$_{80}$; the parameters used in the fits are given in Table \ref{table1}.}
\end{figure}

\begin{figure}
\includegraphics[width=3.2in,height=3in,trim=0 0 0 0]{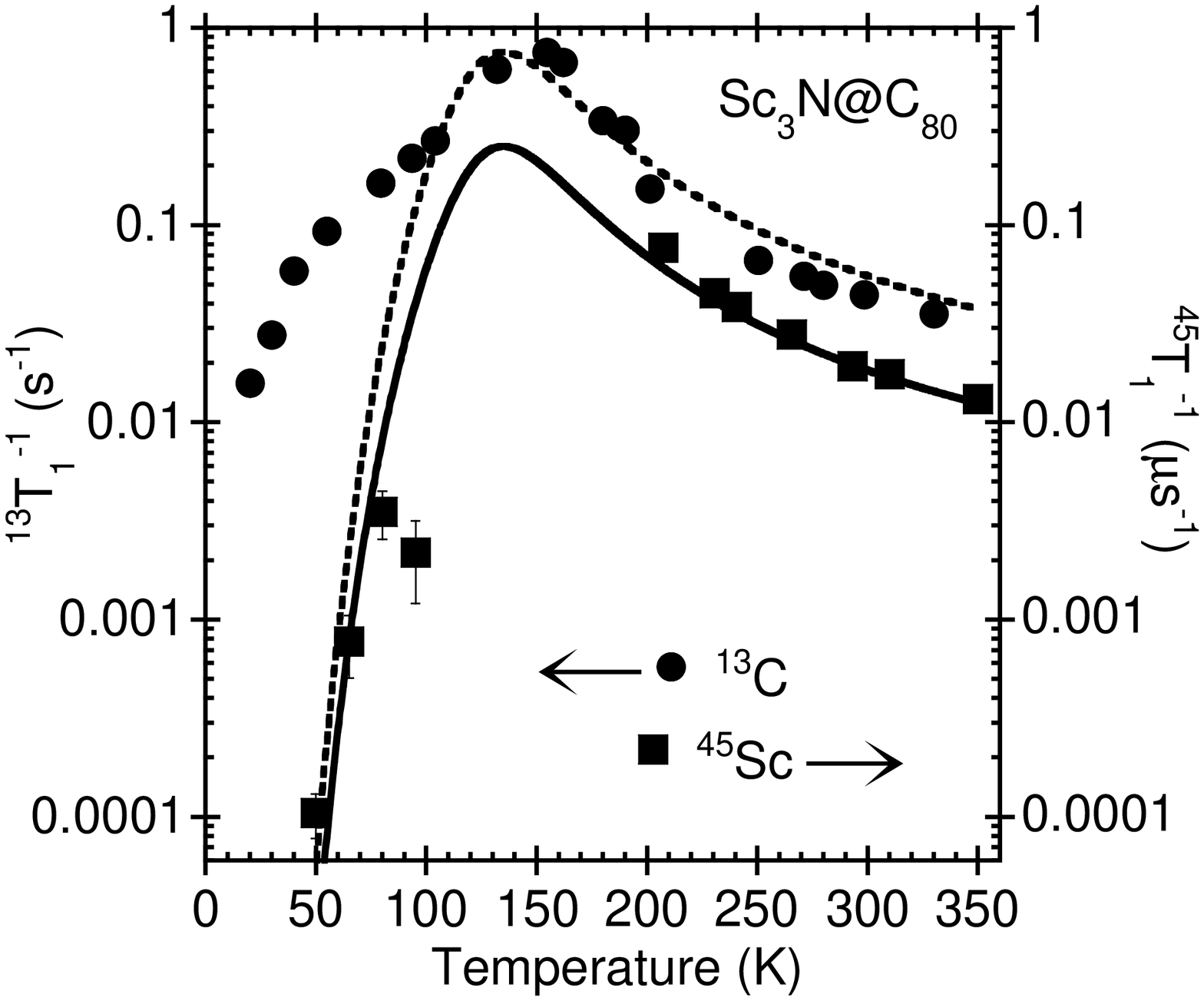}
\caption{\label{c80rate} Spin-lattice relaxation rates 1/T$_{1}$ vs. temperature show similar temperature dependence for both $^{13}$C (circles, left vertical axis) and $^{45}$Sc (squares, right vertical axis) in Sc$_{3}$N@C$_{80}$.  Note the units for 1/T$_{1}$ are s$^{-1}$ for $^{13}$C (left axis) and $\mu$s$^{-1}$ for $^{45}$Sc (right axis).  The lines (dashed for $^{13}$C and solid for $^{45}$Sc) are fits to the data using the model of Tycko \textit{et al.} described in the text that ascribes the relaxation to modulations (due to molecular reorientation) of the chemical shift for $^{13}$C and of the EFG for $^{45}$Sc.  Other than a factor setting the overall scale, the same parameters (given in Table 1) are used to describe the temperature dependence of both sets of data.}
\end{figure}

Measurements of spin-lattice relaxation times (T$_{1}$) for $^{13}$C and $^{45}$Sc also reveal information about molecular dynamics.  Fig.~\ref{carbonrate} plots the $^{13}$C spin-lattice relaxation rate 1/T$_{1}$ versus temperature for Sc$_{3}$N@C$_{80}$, and for comparison we also show the same data for C$_{60}$ as reported by Tycko \textit{et al.}\cite{tycko}  The lines through the data in Fig.~\ref{carbonrate} are fits that we shall describe below.  There is a notable qualitative difference between the T$_{1}$ behaviors of C$_{60}$ and Sc$_{3}$N@C$_{80}$: the C$_{60}$ data show an abrupt change in behavior near 250 K, corresponding to the reported phase transition temperature,\cite{heiney} while the Sc$_{3}$N@C$_{80}$ behavior is relatively smooth throughout the measured temperature range.  Fig.~\ref{c80rate} shows the $^{45}$Sc spin-lattice relaxation rate for Sc$_{3}$N@C$_{80}$ along with that for $^{13}$C; fits to these data (see below) are also included.  Note that the units for the $^{45}$Sc data are $\mu$s$^{-1}$, while those for $^{13}$C are s$^{-1}$:  the $^{45}$Sc relaxation rate is almost six orders of magnitude faster than the $^{13}$C rate.  The reason for this enormous difference is that the $^{45}$Sc relaxation is dominated by EFG fluctuations which are far more effective than the CSA mechanism.

\begin{figure}
\includegraphics[width=3in,height=3in,trim=0 0 0 0]{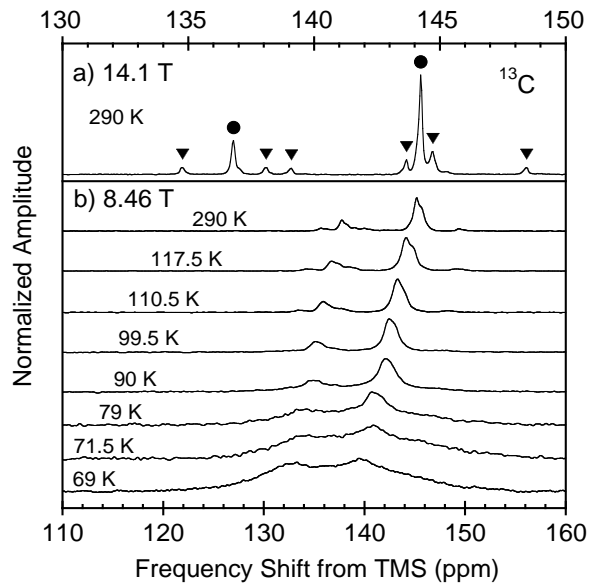}
\caption{\label{mas} Magic Angle Spinning (MAS) NMR spectra are plotted for two different magnetic field strengths and at several temperatures for $^{13}$C in Sc$_{3}$N@C$_{80}$.  The spectra have been normalized, and the line positions are measured in parts-per-million relative to tetramethyl silane (TMS).  In (a), the field strength is 14.1 T and the temperature is 290 K; note the scale is above the axis, running from 130-150 ppm.  The spectrum shows the presence of two isomers of C$_{80}$: one has I$_{h}$ symmetry and two carbon lines with a 3:1 intensity ratio (solid circles), and the other has D$_{5h}$ symmetry with six carbon lines in a 1:1:1:2:2:1 intensity ratio (solid triangles).  In (b), we show the temperature dependence of the spectrum at 8.46 T (lower axis scale, 110-160 ppm).  The primary feature to note is the presence of two lines (due to the I$_{h}$ isomer) over the entire temperature range, indicating the C$_{80}$ cage remains spherically symmetric.  The increase of linewidth with decreasing temperature may indicate a loss of spherical symmetry.}
\end{figure}

To further elucidate the orientational dynamics of the molecules, we measured $^{13}$C spectra using Magic Angle Spinning (MAS) NMR in order to remove the line broadening due to the chemical shift anisotropy at low temperature.  These data are shown in Fig.~\ref{mas}.  Several clarifications are in order here.  First, the MAS linewidths are much smaller than for the motionally-narrowed lines in Fig.~\ref{carbonline}.  The broader lines of Fig.~\ref{carbonline} simply reflect the field inhomogeneity of the magnet used to obtain those data.  Second, one may note that more than two lines are present in the MAS data.  This occurs because as-produced Sc$_{3}$N@C$_{80}$ is a mixture of two isomers: the primary isomer has I$_{h}$ symmetry, yielding two carbon lines, and the second, minor isomer (roughly 10\%) has D$_{5h}$ symmetry, leading to six carbon lines.\cite{duchamp}  The lines associated with each isomer are labeled in Fig.~\ref{mas}(a).  The data of Fig.~\ref{carbonline} do not show the presence of the minor isomer due to the line broadening mentioned previously for those data.  From this point, we shall only focus on the primary isomer with I$_{h}$ symmetry.  Finally, the line positions have a modest temperature dependence of unknown origin.  This temperature dependence partially explains the discrepancy between our NMR data and the previous solution NMR result.

\section{\label{Analysis}Analysis}
To analyze our data quantitatively, we follow the approach Tycko \textit{et al.} used to analyze their $^{13}$C data in C$_{60}$.\cite{tycko}  They used a model in which the spin-lattice relaxation results from fluctuating effective magnetic fields, transverse to the applied field, produced from the anisotropic chemical shift mechanism and modulated in time by molecular reorientations.  They further assumed an exponential autocorrelation function to describe the time dependence of these fluctuations, with a correlation time $\tau$.  Such a model gives the following expression for T$_{1}$:\cite{tycko}
\begin{equation}
 \frac{1}{T_{1}} = \frac{\omega_{1}^{2}\tau}{1 + \omega_{0}^{2}\tau^{2}} ,
\end{equation}
where $\omega_{0}$ is the Larmor frequency, and $\omega_{1}^{2}$ characterizes the mean square strength of the field fluctuations (which we address below).  This expression leads to the well known ÒT$_{1}$ minimumÓ  when $\omega_{0} \tau$ = 1.\cite{cps,bpp}  In terms of these parameters, 1/T$_{1}^{min}$  =  $\omega_{1}^{2}/2\omega_{0}$.   Finally, Tycko \textit{et al.} assumed an activated temperature dependence for $\tau$.  These assumptions lead to the following expression for the spin-lattice relaxation rate:\cite{tycko}
\begin{equation}
 \frac{1}{T_{1}} = \frac{ T_{1}^{min} }  {\cosh(E_{a}/k_{B}T^{min} -  E_{a}/k_{B}T)} ,
\label{equation2}
\end{equation}
where E$_{a}$ is the activation energy, k$_{B}$ is the Boltzmann constant, and T is the temperature.  Tycko \textit{et al.} found that this model could not describe their C$_{60}$ data over the full temperature range.  Instead they required separate sets of parameters T$_{1}^{min}$, T$^{min}$, and E$_{a}$ for the regions above and below 250 K.  The parameters for their fits are listed in Table~\ref{table1}.  Note that in fitting the high-temperature data, they constrained the value of T$_{1}^{min}$ on physical grounds to be equal to the low-temperature value.  The resulting fits are shown in Fig.~\ref{carbonrate}.
\begin{table}
\caption{\label{table1}The values of the parameters T$_{1}^{min}$, T$^{min}$, and E$_{a}$/k$_{B}$ in Eq.~\ref{equation2} are shown for fits to T$_{1}$ data for $^{13}$C in C$_{60}$ (at low and high temperatures from Ref.~\onlinecite{tycko}) and for $^{13}$C and $^{45}$Sc in Sc$_{3}$N@C$_{80}$ (this work).  For C$_{60}$, the high-temperature value of T$_{1}^{min}$ was constrained to be the same as the low-temperature value when fitting the data.  For Sc$_{3}$N@C$_{80}$, the values of T$^{min}$ and E$_{a}$/k$_{B}$ for the $^{45}$Sc data were taken to be the same as the values for the $^{13}$C data, thus leaving only T$_{1}^{min}$ as an adjustable parameter in fitting the $^{45}$Sc data.  The fits are compared to the data in Fig.~\ref{carbonrate} and Fig.~\ref{c80rate}.}
\begin{ruledtabular}
\begin{tabular}{lccc}
& T$_{1}^{min}$ & T$^{min}$ & E$_{a}$/k$_{B}$ \\
\hline
C$_{60}$ (T $<$ 250 K) & 0.806 s & 268 K & 2900 K \\
C$_{60}$ (T $>$ 250 K) & 0.806 s & 74.6 K & 487 K \\
$^{13}$C in Sc$_{3}$N@C$_{80}$ & 1.33 s & 135 K & 811 K \\
$^{45}$Sc in Sc$_{3}$N@C$_{80}$ & 4.0 $\mu$s & 135 K & 811 K \\
\end{tabular}
\end{ruledtabular}
\end{table}

In contrast to C$_{60}$, our T$_{1}$ results for Sc$_{3}$N@C$_{80}$ follow reasonably well a single pattern of activated behavior over the full temperature range.  Also shown in Fig.~\ref{carbonrate} is a fit to our $^{13}$C data using Eq.~\ref{equation2}, with the parameters given in Table~\ref{table1}.  The activation energy 811 K is not too far from the high temperature phase activation energy (487 K) found for C$_{60}$, suggesting behavior more nearly like the high temperature ``rotator'' phase.  At lower temperatures the $^{13}$C spin-lattice relaxation rate systematically exceeds the fit, perhaps suggesting the emergence of some other relaxation mechanism.  At our resonance frequency $\omega_{0}/2\pi$ = 94 MHz and using the values of T$_{1}^{min}$ = 1.33 s and T$^{min}$ = 135 K from the fit, we are able to estimate a correlation time of 1.7 ns at that temperature and 62 ps at T = 300 K, some 5 times longer than the corresponding estimate for C$_{60}$.\cite{tycko}

We can utilize our chemical shift data to provide a consistency check for the model by estimating T$_{1}^{min}$ from the expression 1/T$_{1}^{min}$ = $\omega_{1}^{2}/2\omega_{0}$.  Johnson \textit{et al.}\cite{johnson} showed that for the assumed CSA relaxation mechanism, the phenomenological parameter $\omega_{1}^{2} = 2\omega_{0}^{2}S^{2}/15$, with $S^{2} \equiv \Delta \sigma^{2}(1 + \eta^{2}/3)$, $\Delta \sigma \equiv \sigma_{33} - (\sigma_{22} + \sigma_{11})/2$, and $\eta \equiv (\sigma_{22} - \sigma_{11})/(2 \Delta \sigma /3)$.  The $\sigma_{ii}$ are the chemical shift tensor components with $\sigma_{33} > \sigma_{22} > \sigma_{11}$.  Our shift values give $S^{2} \approx 3$ x $10^{-8}$, and from this we estimate T$_{1}^{min} \approx$ 0.85 s.  This value is comparable to the experimentally measured number, although it is unclear why this estimate is shorter than the measured value.

We have also applied the model of Tycko \textit{et al.} to our $^{45}$Sc data, although in this case the relaxation is due to fluctuations of the electric field gradient brought about by molecular reorientations.  Like $^{13}$C, the scandium relaxation rate reaches a T$_{1}^{min}$ at temperatures between 100 K and 200 K, although for $^{45}$Sc, T$_{1}^{min}$ is too small to be measured.  Overall there is impressive similarity between the $^{13}$C and $^{45}$Sc data.  Two fits are also given in Fig.~\ref{c80rate}.  For the $^{13}$C data we apply a fit identical to that shown in Fig.~\ref{carbonrate}, based on Eq.~\ref{equation2} and employing the parameters in Table~\ref{table1}.  In order to emphasize the similarity of the $^{45}$Sc and $^{13}$C data, we also fit the $^{45}$Sc data to the form of Eq.~\ref{equation2}, using the values for E$_{a}$/k$_{B}$ and T$^{min}$ taken from the $^{13}$C fit and allowing only T$_{1}^{min}$ as an adjustable parameter.  Note that the resonance frequencies of $^{45}$Sc and $^{13}$C are only $\sim 2\%$ different.  The resulting fit tracks the $^{45}$Sc data remarkably well with T$_{1}^{min}$ = 4.0 $\mu$s.

As a check on our approach to the $^{45}$Sc data, we consider the magnitude of the relaxation rate.  For a spin-I nucleus experiencing EFG fluctuations in the fast correlation time limit appropriate for the room temperature data well above T$^{min}$, the relaxation for isotropic reorientational motion with correlation time $\tau$ (with asymmetry parameter $\eta = 0$) is given by:\cite{abragam}  
\begin{equation}
 \frac{1}{T_{1}} = \frac{3}{40} \frac{2I + 3}{I^{2} (2I - 1)} \left(\frac{e^{2}qQ}{\hbar}\right)^{2}  \tau .
\end{equation}
From the room temperature correlation time estimated above as 62 ps and the measured e$^{2}$qQ/h = 67.9 MHz, the calculated rate is 0.12 $\mu$s$^{-1}$, some seven times the measured rate of 0.018 $\mu$s$^{-1}$.  This discrepancy may indicate the motion of the Sc$_{3}$N is restricted rather than isotropic.

\section{\label{Bound}Is the S\lowercase{c}$_{3}$N bound to the C$_{80}$ cage?}
The remaining issue is the bonding relation between the C$_{80}$ cage and the Sc$_{3}$N inside.  In principle, our data should be able to address this since we have one probe ($^{13}$C) of the C$_{80}$ cage and another probe ($^{45}$Sc) of the Sc$_{3}$N.  However, no definitive and satisfying answer is forthcoming.

The previous experimental and theoretical results can be summarized as follows.  Raman/IR experiments show additional phonons that point to a reduced symmetry for the Sc$_{3}$N@C$_{80}$ molecule at both 80 K and 300 K, suggesting the Sc$_{3}$N is fixed to the C$_{80}$ cage.\cite{krause}  The solution NMR observation\cite{steve} of two carbon lines with a 3:1 intensity ratio requires a C$_{80}$ cage with I$_{h}$ symmetry: that result precisely matches what one expects for C$_{80}$ with I$_{h}$ symmetry, since that cage has 60 corannulene carbons (at the junction of a five-membered ring and two six-membered rings) and 20 pyrene carbons (at the junction of three six-membered rings).  The other isomers of C$_{80}$, being of lower symmetry, have more than two distinct carbon sites, which would yield carbon spectra with more than two lines and intensity ratios of no more than 2:1 for any set of lines.\cite{sun}  Theoretical calculations for the energy barrier for motion of the Sc$_{3}$N inside the C$_{80}$ cage found a value of 0.144 eV, yielding a characteristic time scale for the motion of 1.53 ps at 575 K.\cite{vietze}  Based on their calculations and on these two experimental results, Vietze and Seifert ascribed the discrepancy between the solution NMR and Raman/IR results to the time scales for the two experiments.  According to this explanation, since the typical Raman/IR time scale $\sim 0.1$ ps is faster than the Sc$_{3}$N motion, that experiment saw the instantaneous position of the Sc$_{3}$N and thus a lack of spherical symmetry.  Because the solution NMR time scale $\sim 1$ ns is much slower than the Sc$_{3}$N motion, that experiment only found an average position for the Sc$_{3}$N, thus creating the appearance of I$_{h}$ symmetry.

We consider the data reported here in the context of this picture.  The $^{13}$C spectra (both MAS and non-MAS) consisting of two lines in a 3:1 ratio can only derive from a C$_{80}$ cage with I$_{h}$ symmetry, meaning the Sc$_{3}$N cannot be fixed to the C$_{80}$ cage on the NMR time scale.  This is consistent with the above ideas.  We would then interpret the line broadening observed at low temperature by MAS as the loss of spherical symmetry of the Sc$_{3}$N@C$_{80}$ system due to the cessation of motion of the Sc$_{3}$N. The $^{13}$C and $^{45}$Sc relaxation rates both have similar temperature behavior, each displaying T$_{1}$ minima between 100 K and 200 K, and in fact we can fit both sets of data with identical activation energies.  This would most simply be explained by a single physical mechanism as the cause of the motion of both the C$_{80}$ cage and the Sc$_{3}$N, which would seem to require they be bound together.  If, however, we require the C$_{80}$ cage and the Sc$_{3}$N not be bound together, we must then conclude that the similar T$_{1}$ behavior is sheer coincidence.  While we cannot rule this out, it would seem to require a remarkable set of circumstances to conspire to bring this about.  To argue the opposite position, that the C$_{80}$ cage and the Sc$_{3}$N are fixed together, one must overcome the symmetry breaking effect of the bonding of the Sc$_{3}$N to the C$_{80}$ cage:  the loss of spherical symmetry would lead to more than two carbon lines, contrary to our observation.  The line broadening observed in the MAS data may be indicating such a loss of symmetry, but that remains speculative.

\begin{acknowledgments}
J. A. Martindale thanks P. C. Hammel, P. Wigen, and M. Boss for beneficial discussions.  S. Stevenson acknowledges funding from the NSF EPSCOR and NSF CAREER Grant CHE 0547988.  I. Heinmaa and R. Stern acknowledge support from the Estonian Science Foundation.
\end{acknowledgments}

\bibliography{paper.bib}

\begin{thebibliography}{15}
\expandafter\ifx\csname natexlab\endcsname\relax\def\natexlab#1{#1}\fi
\expandafter\ifx\csname bibnamefont\endcsname\relax
  \def\bibnamefont#1{#1}\fi
\expandafter\ifx\csname bibfnamefont\endcsname\relax
  \def\bibfnamefont#1{#1}\fi
\expandafter\ifx\csname citenamefont\endcsname\relax
  \def\citenamefont#1{#1}\fi
\expandafter\ifx\csname url\endcsname\relax
  \def\url#1{\texttt{#1}}\fi
\expandafter\ifx\csname urlprefix\endcsname\relax\def\urlprefix{URL }\fi
\providecommand{\bibinfo}[2]{#2}
\providecommand{\eprint}[2][]{\url{#2}}

\bibitem[{\citenamefont{Stevenson et~al.}(1999)\citenamefont{Stevenson, Rice,
  Glass, Harich, Cromer, Jordan, Craft, Hadju, Bible, Olmstead et~al.}}]{steve}
\bibinfo{author}{\bibfnamefont{S.}~\bibnamefont{Stevenson}},
  \bibinfo{author}{\bibfnamefont{G.}~\bibnamefont{Rice}},
  \bibinfo{author}{\bibfnamefont{T.}~\bibnamefont{Glass}},
  \bibinfo{author}{\bibfnamefont{K.}~\bibnamefont{Harich}},
  \bibinfo{author}{\bibfnamefont{F.}~\bibnamefont{Cromer}},
  \bibinfo{author}{\bibfnamefont{M.~R.} \bibnamefont{Jordan}},
  \bibinfo{author}{\bibfnamefont{J.}~\bibnamefont{Craft}},
  \bibinfo{author}{\bibfnamefont{E.}~\bibnamefont{Hadju}},
  \bibinfo{author}{\bibfnamefont{R.}~\bibnamefont{Bible}},
  \bibinfo{author}{\bibfnamefont{M.~M.} \bibnamefont{Olmstead}},
  \bibnamefont{et~al.}, \bibinfo{journal}{Nature}
  \textbf{\bibinfo{volume}{401}}, \bibinfo{pages}{55} (\bibinfo{year}{1999}).

\bibitem[{\citenamefont{Heiney et~al.}(1991)\citenamefont{Heiney, Fischer,
  McGhie, Romanow, Denenstein, McCauley, and Smith~III}}]{heiney}
\bibinfo{author}{\bibfnamefont{P.~A.} \bibnamefont{Heiney}},
  \bibinfo{author}{\bibfnamefont{J.~E.} \bibnamefont{Fischer}},
  \bibinfo{author}{\bibfnamefont{A.~R.} \bibnamefont{McGhie}},
  \bibinfo{author}{\bibfnamefont{W.~J.} \bibnamefont{Romanow}},
  \bibinfo{author}{\bibfnamefont{A.~M.} \bibnamefont{Denenstein}},
  \bibinfo{author}{\bibfnamefont{J.~P.} \bibnamefont{McCauley},
  \bibfnamefont{Jr.}}, \bibnamefont{and} \bibinfo{author}{\bibfnamefont{A.~B.}
  \bibnamefont{Smith~III}}, \bibinfo{journal}{Phys.\ Rev.\ Lett.}
  \textbf{\bibinfo{volume}{66}}, \bibinfo{pages}{2911} (\bibinfo{year}{1991}).

\bibitem[{\citenamefont{Tycko et~al.}(1991)\citenamefont{Tycko, Dabbagh,
  Fleming, Haddon, Makhija, and Zahurak}}]{tycko}
\bibinfo{author}{\bibfnamefont{R.}~\bibnamefont{Tycko}},
  \bibinfo{author}{\bibfnamefont{G.}~\bibnamefont{Dabbagh}},
  \bibinfo{author}{\bibfnamefont{R.~M.} \bibnamefont{Fleming}},
  \bibinfo{author}{\bibfnamefont{R.~C.} \bibnamefont{Haddon}},
  \bibinfo{author}{\bibfnamefont{A.~V.} \bibnamefont{Makhija}},
  \bibnamefont{and} \bibinfo{author}{\bibfnamefont{S.~M.}
  \bibnamefont{Zahurak}}, \bibinfo{journal}{Phys.\ Rev.\ Lett.}
  \textbf{\bibinfo{volume}{67}}, \bibinfo{pages}{1886} (\bibinfo{year}{1991}).

\bibitem[{\citenamefont{Johnson et~al.}(1992)\citenamefont{Johnson, Yannoni,
  Dorn, Salem, and Bethune}}]{johnson}
\bibinfo{author}{\bibfnamefont{R.~D.} \bibnamefont{Johnson}},
  \bibinfo{author}{\bibfnamefont{C.~S.} \bibnamefont{Yannoni}},
  \bibinfo{author}{\bibfnamefont{H.~C.} \bibnamefont{Dorn}},
  \bibinfo{author}{\bibfnamefont{J.~R.} \bibnamefont{Salem}}, \bibnamefont{and}
  \bibinfo{author}{\bibfnamefont{D.~S.} \bibnamefont{Bethune}},
  \bibinfo{journal}{Science} \textbf{\bibinfo{volume}{255}},
  \bibinfo{pages}{1235} (\bibinfo{year}{1992}).

\bibitem[{\citenamefont{Pennington and Stenger}(1996)}]{review}
\bibinfo{author}{\bibfnamefont{C.~H.} \bibnamefont{Pennington}}
  \bibnamefont{and} \bibinfo{author}{\bibfnamefont{V.~A.}
  \bibnamefont{Stenger}}, \bibinfo{journal}{Rev.\ Mod.\ Phys.}
  \textbf{\bibinfo{volume}{68}}, \bibinfo{pages}{855} (\bibinfo{year}{1996}).

\bibitem[{\citenamefont{Samoson et~al.}(2004)\citenamefont{Samoson, Tuherm,
  Past, Reinhold, Anupold, and Heinmaa}}]{stern}
\bibinfo{author}{\bibfnamefont{A.}~\bibnamefont{Samoson}},
  \bibinfo{author}{\bibfnamefont{T.}~\bibnamefont{Tuherm}},
  \bibinfo{author}{\bibfnamefont{J.}~\bibnamefont{Past}},
  \bibinfo{author}{\bibfnamefont{A.}~\bibnamefont{Reinhold}},
  \bibinfo{author}{\bibfnamefont{T.}~\bibnamefont{Anupold}}, \bibnamefont{and}
  \bibinfo{author}{\bibfnamefont{I.}~\bibnamefont{Heinmaa}},
  \bibinfo{journal}{Top.\ Curr.\ Chem.} \textbf{\bibinfo{volume}{246}},
  \bibinfo{pages}{15} (\bibinfo{year}{2004}).

\bibitem[{\citenamefont{Yannoni et~al.}(1991)\citenamefont{Yannoni, Johnson,
  Meijer, Bethune, and Salem}}]{yannoni}
\bibinfo{author}{\bibfnamefont{C.~S.} \bibnamefont{Yannoni}},
  \bibinfo{author}{\bibfnamefont{R.~D.} \bibnamefont{Johnson}},
  \bibinfo{author}{\bibfnamefont{G.}~\bibnamefont{Meijer}},
  \bibinfo{author}{\bibfnamefont{D.~S.} \bibnamefont{Bethune}},
  \bibnamefont{and} \bibinfo{author}{\bibfnamefont{J.~R.} \bibnamefont{Salem}},
  \bibinfo{journal}{J.\ Phys.\ Chem.} \textbf{\bibinfo{volume}{95}},
  \bibinfo{pages}{9} (\bibinfo{year}{1991}).

\bibitem[{\citenamefont{Slichter}(1992)}]{cps}
\bibinfo{author}{\bibfnamefont{C.~P.} \bibnamefont{Slichter}},
  \emph{\bibinfo{title}{Principles of Magnetic Resonance}}
  (\bibinfo{publisher}{Springer-Verlag}, \bibinfo{address}{New York},
  \bibinfo{year}{1992}).

\bibitem[{\citenamefont{Cohen and Reif}(1957)}]{cohenreif}
\bibinfo{author}{\bibfnamefont{M.~H.} \bibnamefont{Cohen}} \bibnamefont{and}
  \bibinfo{author}{\bibfnamefont{F.}~\bibnamefont{Reif}}, in
  \emph{\bibinfo{booktitle}{Solid State Physics}}, edited by
  \bibinfo{editor}{\bibfnamefont{F.}~\bibnamefont{Seitz}} \bibnamefont{and}
  \bibinfo{editor}{\bibfnamefont{D.}~\bibnamefont{Turnbull}}
  (\bibinfo{publisher}{Academic Press}, \bibinfo{address}{New York},
  \bibinfo{year}{1957}), p. \bibinfo{pages}{321}.

\bibitem[{\citenamefont{Duchamp et~al.}(2003)\citenamefont{Duchamp, Demortier,
  Fletcher, Dorn, Iezzi, Glass, and Dorn}}]{duchamp}
\bibinfo{author}{\bibfnamefont{J.~C.} \bibnamefont{Duchamp}},
  \bibinfo{author}{\bibfnamefont{A.}~\bibnamefont{Demortier}},
  \bibinfo{author}{\bibfnamefont{K.~R.} \bibnamefont{Fletcher}},
  \bibinfo{author}{\bibfnamefont{D.}~\bibnamefont{Dorn}},
  \bibinfo{author}{\bibfnamefont{E.~B.} \bibnamefont{Iezzi}},
  \bibinfo{author}{\bibfnamefont{T.}~\bibnamefont{Glass}}, \bibnamefont{and}
  \bibinfo{author}{\bibfnamefont{H.~C.} \bibnamefont{Dorn}},
  \bibinfo{journal}{Chem.\ Phys.\ Lett.} \textbf{\bibinfo{volume}{375}},
  \bibinfo{pages}{655} (\bibinfo{year}{2003}).

\bibitem[{\citenamefont{Bloembergen et~al.}(1948)\citenamefont{Bloembergen,
  Purcell, and Pound}}]{bpp}
\bibinfo{author}{\bibfnamefont{N.}~\bibnamefont{Bloembergen}},
  \bibinfo{author}{\bibfnamefont{E.~M.} \bibnamefont{Purcell}},
  \bibnamefont{and} \bibinfo{author}{\bibfnamefont{R.~V.} \bibnamefont{Pound}},
  \bibinfo{journal}{Phys.\ Rev.} \textbf{\bibinfo{volume}{73}},
  \bibinfo{pages}{679} (\bibinfo{year}{1948}).

\bibitem[{\citenamefont{Abragam}(1961)}]{abragam}
\bibinfo{author}{\bibfnamefont{A.}~\bibnamefont{Abragam}},
  \emph{\bibinfo{title}{Principles of Nuclear Magnetism}}
  (\bibinfo{publisher}{Oxford University Press}, \bibinfo{address}{New York},
  \bibinfo{year}{1961}), p. \bibinfo{pages}{314}.

\bibitem[{\citenamefont{Krause et~al.}(2001)\citenamefont{Krause, Kuzmany,
  Georgi, Dunsch, Vietze, and Seifert}}]{krause}
\bibinfo{author}{\bibfnamefont{M.}~\bibnamefont{Krause}},
  \bibinfo{author}{\bibfnamefont{H.}~\bibnamefont{Kuzmany}},
  \bibinfo{author}{\bibfnamefont{P.}~\bibnamefont{Georgi}},
  \bibinfo{author}{\bibfnamefont{L.}~\bibnamefont{Dunsch}},
  \bibinfo{author}{\bibfnamefont{K.}~\bibnamefont{Vietze}}, \bibnamefont{and}
  \bibinfo{author}{\bibfnamefont{G.}~\bibnamefont{Seifert}},
  \bibinfo{journal}{J.\ Chem.\ Phys.} \textbf{\bibinfo{volume}{115}},
  \bibinfo{pages}{6596} (\bibinfo{year}{2001}).

\bibitem[{\citenamefont{Sun and Kertesz}(2000)}]{sun}
\bibinfo{author}{\bibfnamefont{G.}~\bibnamefont{Sun}} \bibnamefont{and}
  \bibinfo{author}{\bibfnamefont{M.}~\bibnamefont{Kertesz}},
  \bibinfo{journal}{Chem.\ Phys.\ Lett.} \textbf{\bibinfo{volume}{328}},
  \bibinfo{pages}{387} (\bibinfo{year}{2000}).

\bibitem[{\citenamefont{Vietze and Seifert}(2002)}]{vietze}
\bibinfo{author}{\bibfnamefont{K.}~\bibnamefont{Vietze}} \bibnamefont{and}
  \bibinfo{author}{\bibfnamefont{G.}~\bibnamefont{Seifert}}, in
  \emph{\bibinfo{booktitle}{Structural and Electronic Properties of Molecular
  Nanostructures}}, edited by
  \bibinfo{editor}{\bibfnamefont{H.}~\bibnamefont{Kuzmany}},
  \bibinfo{editor}{\bibfnamefont{G.}~\bibnamefont{Fink}}, \bibnamefont{and}
  \bibinfo{editor}{\bibfnamefont{M.}~\bibnamefont{Mehring}}
  (\bibinfo{publisher}{Springer}, \bibinfo{address}{Secaucus, NJ},
  \bibinfo{year}{2002}), p.~\bibinfo{pages}{39}.

\end{thebibliography}

\end{document}